\documentclass[journal,onecolumn,12pt]{IEEEtran}
\usepackage{booktabs} % For formal tables

\usepackage{lineno}
\usepackage{color}
\usepackage{multirow}
\usepackage{array}
\usepackage{amsmath}
\usepackage{balance}
\usepackage{wrapfig}
\usepackage[justification=centering]{caption} %centering the caption of tables & figures
\usepackage{tikz}
\usepackage{amssymb}
\newcommand*{\circled}[1]{\lower.7ex\hbox{\tikz\draw (0pt, 0pt)%
    circle (.5em) node {\makebox[1em][c]{\small #1}};}}
\usepackage{indentfirst}  % set the indent for the first paragraph
\newcommand{\PreserveBackslash}[1]{\let\temp=\\#1\let\\=\temp}
\newcolumntype{C}[1]{>{\PreserveBackslash\centering}p{#1}}
\newcolumntype{R}[1]{>{\PreserveBackslash\raggedleft}p{#1}}
\newcolumntype{L}[1]{>{\PreserveBackslash\raggedright}p{#1}}
\renewcommand{\thefootnote}{\fnsymbol{footnote}} %将脚注符号设置为fnsymbol类型，即特殊符号表示, 此外还有数字式样为\fnsymbol,\roman,\Roman,\alph,\ALph,\arabic
\newcommand{\tabincell}[2]{\begin{tabular}{@{}#1@{}}#2\end{tabular}}

\newtheorem{myDef}{Definition}

\newtheorem{myTheo}{Theorem}
\newtheorem{myLemma}{Lemma}[myTheo]

\makeatletter
\newcommand\sixteen{\@setfontsize\sixteen{17pt}{6}}
\renewcommand{\maketitle}{\bgroup\setlength{\parindent}{0pt}
\begin{flushleft}
\sixteen\bfseries \@title
\medskip
\end{flushleft}
{\@author}
\egroup
}
\makeatother

\usepackage{setspace}
\usepackage{listings} %for codes

\usepackage{cite}

\ifCLASSINFOpdf
\else
\fi

\begin{document}
%\linenumbers

\renewcommand{\baselinestretch}{1.1}
\title{TinyAKE: A More Practicable and Trustable Scheme for\\ \vspace{0.2em}Authenticated Key Establishment in WSNs}

\author{Fajun Sun\textsuperscript{1}, Selena He\textsuperscript{2}, Xiaotong Zhang\textsuperscript{1}, Fanfan Shen\textsuperscript{3}, Qingan Li\textsuperscript{1}, Yanxiang He\textsuperscript{1}\footnotemark[2]\\\\
\small{\textsuperscript{1}\emph{School of Computer Science, Wuhan University, Wuhan 430072, China}}\\
\small{\textsuperscript{2}\emph{Department of Computer Science, Kennesaw State University, Marietta 30060, USA}}\\
\small{\textsuperscript{3}\emph{School of Information Engineering, Nanjing Audit University, Nanjing 211815, China}}\\\\
\vspace{2em}\footnotemark[2]Correspondence should be addressed to Yanxiang He; yxhe@whu.edu.cn
}

%Fajun Sun, sunfajun@whu.edu.cn \orcid{0000-0002-8803-9801}
%Selena He, she4@kennesaw.edu
%Fanfan Shen,ffshen@whu.edu.cn
%Qingan Li,qingan@whu.edu.cn
%Yanxiang He, yxhe@whu.edu.cn \orcid{0000-0002-8648-993X}

\maketitle

\begin{abstract}
\noindent The characteristics of high loss rate, resource constraint, being eager for good security haven't been fully considered in the existing key establishment protocols of wireless sensor networks. Analyzing the key establishing problem from the MAC and physical layers, existing protocols are not practicable enough due to their overlong agreement packets and single round key establishment. To mitigate the impact of these problems, a group of design principles for secure sensor networks has been presented and TinyAKE, an authenticated key transport protocol based on lightweight certificate, is proposed in this paper. The security of TinyAKE are proved with the theory of indistinguishability, meanwhile, the correctness is also proved, the performance is analyzed and compared with the existing similar protocols. Finally TinyAKE is implemented in the TinyOS with TinyECC. Our evaluation shows that TinyAKE is a more practicable and trustable authenticated key establishment protocol than existing protocols. The experimental result shows that the key transport with certificate mechanism is feasible in WSNs. Moreover, the simulation results show that the optimal number of repeated negotiation is one when the secure connectivity rate of TinyAKE is improved by using the repeated key negotiation.%Moreover, the most cost-effective number of retransmission is one when security connectivity of TinyAKE is improved by repeated key agreement indicating by the simulation results implemented in the TinyOS. %
\end{abstract}


\IEEEpeerreviewmaketitle

\section{Introduction}\label{ch:Intro}
In the past 20 years of research on WSNs (Wireless Sensor Networks), the security of WSNs has always been a hot topic \cite{Shim2016A}, even after the coming of fog computing \cite{Wang2019A}, edge computing \cite{Cai2019Data} and Internet of Things \cite{Saeed2019AKAIoTs}. Because of resource limitation and openness of WSNs \cite{Carman2000Constraints}, whether to use asymmetric cryptography to improve security or to use symmetric cryptography to save resources has even been a controversial focus in security researches of WSNs.
%the application of asymmetric cryptography with high security or the use of symmetric cryptography to save resources has been a controversial issue in security research.
In earlier years, researchers focused on the applications of symmetric key mechanism \cite{Perrig2002SPINS,Karlof2004TinySec,Eschenauer2002A,Chan2003Random}, but it has inherent shortcomings in key exchange. After the feasibility has been approved \cite{Gura2004Comparing,Malan2004A,Watro2004TinyPK}, the public key mechanism has attracted a lot of attentions and has been widely studied in WSNs. 
% in order to make up for the deficiency of symmetric key mechanism, after a few years of research on the symmetric key mechanism, the feasibility of public key mechanism in WSNs attracts considerable attention and makes the public key mechanism of WSNs be extensively studied.
From the view of existing researches, in the WSNs with limited resources, public key mechanism can also be applied when needed, as is even more feasible in the rechargeable WSNs \cite{Zameni2020Two}. The existing researches on public key mechanism in WSNs mainly focus on the following three aspects \cite{He2019A}: primitives, key management (especially the establishment of pairwise keys between nodes in the network, i.e. key establishment), authentication and access control (especially the authentication of users outside the WSNs). Most researchers \cite{Wander2005Energy,Shim2016A} believe that the public key mechanism should only be used for key establishment (abbreviated as KE), which can not only build the foundation of network security, but also avoid deriving excessive expenses and attacks from the overuse of public key mechanism. Since the establishment of security links is the foundation of all the other network functions, in this paper, we mainly focuses on the discussion of KE, especially authenticated KE (abbreviated as AKE). The existing researches on KE are mainly divided into two categories: unauthenticated KE (abbreviated as UKE) schemes and AKE schemes, where AKE schemes in WSNs  are divided into four sub-categories according to the way of authentication: certificate-based \cite{Watro2004TinyPK,Wander2005Energy}, hash-based \cite{Qin2014An,Nadir2016Establishing}, certificateless \cite{Arazi2007A,Wang2011Public,Seo2015Effective,Saeed2019AKAIoTs} and other schemes \cite{Malan2008Implementing,Tufail2015A}. 
However most of the existing literature \cite{Arazi2007A,Wang2011Public,Seo2015Effective,Saeed2019AKAIoTs} are based on the assumption of reliable transport layer and prefer certificateless key agreement. We think there are at least three issues that need to be addressed:

$\bullet$ Among the existing standards for WSNs, we should follow IEEE802.15.4 \cite{Ieee2012IEEE} because it is adopted by most WSN protocol stacks, such as Zigbee and 6LoWPAN. The length of agreement frame should be designed to not exceed its allowable maximum (e.g. aMaxPHYPacketSize: 127B)\cite{Sciancalepore2017Public}. Otherwise, the packet needs to be split when it is transmitted to the lower layer. In this case, an attacker can destroy the transmission of an entire packet by simply destroying the communication of any one fragment, thus reducing security. Consequently, it is worth deliberating how to control the packet length in the range of MTU (Maximum Transmission Unit) and meet the security requirements in the resource-constrained WSNs.

$\bullet$ There are high loss possibility of key establishment packets under the bad link condition of WSNs. How can we design the suitable control mechanism on reliability, so as to improve the coverage rate of security links? 

$\bullet$ Previous researches \cite{Malan2004A,Wander2005Energy,Seo2015Effective,Sciancalepore2017Public,Saeed2019AKAIoTs} mainly focused on the key agreement based on Diffie-Hellman protocol \cite{Diffie1976New}, but there is little analysis and research \cite{Nadir2016Establishing} on whether the key transport mechanism with light certificate is feasible in WSNs. Perhaps certificate-based approaches can solve the problem of overlong packages.

As a result, we focus on the above three issues to discuss the design of a practicable and trustable AKE scheme in this paper. Our contributions are summarized as follows.

\textbf{Discovering flaws and design principles}: We point out some design flaws in the existing security protocols of WSNs, such as overlong frame and authentication flaws, etc. The main reason for these flaws is that the design is not considered from the MAC and physical layers. So we reemphasize the view of cross layer security design, and add some design principles on the basis of existing researches \cite{Law2003An} to mitigate these flaws. 

\textbf{A novel AKE protocol based on WSNs}: Then we incorporate the above ideas into the design and implementation of TinyAKE (Tiny Authenticated Key Establishment protocol), including the constraint of packet length, authentication, and retransmission of key agreement packet etc. %As far as we know, the key establishment protocol with Three-Authentication for WSNs has not been proposed elsewhere. Compared with existing protocols, the most significant feature of TinyAKE is its authentication, which makes it securer than other protocols at reasonable cost. As Wander \cite{Wander2005Energy} and other scholars have discussed, this is feasible for infrequent key establishment operations in WSNs. 
To the best of our knowledge, this is the first to integrate security and reliability (mainly means retransmission)  simultaneously in AKE protocols of WSNs. We see TinyAKE as a prototype scheme of trustable WSNs.

\textbf{Experimental researches on reliable AKE}: We carried out the test of packet length and the simulation on retransmission of key establishment packets. We preliminarily discussed how to improve the security connectivity rate of the bad communication network via packet retransmission mechanism. This is the first experimental research on retransmission of agreement packets in a public key-based AKE protocol.

The remainder of this paper is organized as follows. Section \ref{ch:RW} describes the related work. Section \ref{ch:Pre} introduces the background information and preliminaries of TinyAKE. Section \ref{ch:DAKE} presents the protocol TinyAKE and related design principles. Section \ref{ch:PE} gives the proof and performance evaluation of TinyAKE. Section \ref{ch:EA} describes the experiment on TelosB and simulation on MICAz. The experimental and simulating results of TinyAKE are also analyzed in this section. At last, section \ref{ch:Con} concludes this paper.

\section{Related Work}\label{ch:RW}
There are some approaches have been proposed based on PKC (Public Key Cryptography) to support key establishment in WSNs. Earlier in 2004, Gura et al. \cite{Gura2004Comparing} from Sun Microsystems Laboratories implemented elliptic curve point multiplication for 160-bit, 192-bit, and 224-bit NIST/SECG curves over $GF(p)$ and RSA-1024 and RSA-2048 on two 8-bit micro-controllers. Their work breaks the idea that public keys are too expensive to use on small devices. They proposed an algorithm to reduce the number of memory accesses. Their research shows that public-key cryptography is viable on small devices without hardware acceleration, and compared with RSA, the advantage of ECC is remarkable, especially in terms of speed and storage. Then Malan et al. \cite{Malan2004A} presented the first implementation (we call it extended PKI, abbreviated as EPKI) of elliptic curve cryptography for WSNs based on the MICA2 mote. Their analysis shows that public-key infrastructure is viable for symmetric keys' distribution on the MICA2. But their key exchange protocol is based on pure Diffie-Hellman(DH) in which some security flaws have been found \cite{Law2003An,Wang2011Public}. The main problem is, due to the potential Man-In-The-Middle (MITM) attack, DH cannot be directly used in WSNs \cite{Wang2011Public}. Subsequently the ECC-based LSSL (Lightweight Secure Sockets Layer) in \cite{Wander2005Energy} solved this problem by the authentication of public key with certificate, and applied ECDH to implement the key exchange. But their shared pairwise key is static, that is, the pairwise key of a pair of nodes is a fixed value, so once the key is broken, the session security will be lost forever. TinyPK \cite{Watro2004TinyPK} and RSA-based LSSL have nearly implemented ephemeral key with the required authentications by key transport, but unfortunately, they are based on the RSA primitives which has been recognized as impracticable in WSNs \cite{Shim2016A,Carman2000Constraints,Gura2004Comparing}. At the same time, Carman \cite{Carman2000Constraints} and Wander et al. \cite{Wander2005Energy} pointed out that the cost of public key operations was enormous, a scalar multiplication takes about 0.81 seconds while a RSA-1024 private-key modular exponentiation takes nearly 10.99 seconds \cite{Gura2004Comparing}. For both RSA-1024 and ECC-160, the public-key computation dominates consuming 82\% and 72\% of the energy respectively, and communication costs are second \cite{Wander2005Energy}. Therefore, if we employ the public key scheme, we should apply it in an infrequent operation, e.g. a handshake of a key establishment. Recently, Nadir et al. \cite{Nadir2016Establishing} presented a protocol (abbreviated as NZMA) with which we can establish symmetric pairwise-keys using public key cryptography. Just as Du et al. \cite{Du2005An} do, they use the hash function to speed the public key authentication and reduce energy consumption. Comparing the Merkle tree of Du et al. \cite{Du2005An}, NZMA only uses a table to store the hash values of all the nodes, which can reduce communication traffic but also increase the requiring of storage. And the size of NZMA's hash table is proportional to the number of nodes, so it lost the scalability as well. Most importantly a captured node can expose the hash table of whole nodes, which may lead to a Hash Collision attack \cite{Wang2005Finding} on the whole network. With the increasing attention of AKA schemes based on certificateless PKC (CL-PKC), Seo et al. \cite{Seo2015Effective} proposed a certificateless key management scheme CL-EKM based on dynamic HWSN (Hierarchical Wireless Sensor Networks), and Mutaz et al. proposed an authenticated key agreement protocol, named AKAIoTs \cite{Saeed2019AKAIoTs}, for sharing key between cloud server and sensors in Internet of Things (IoT). Both of them are based on the certificateless key agreement technology, but their packets exceed the maximum allowed payload of the MAC frame defined by IEEE802.15.4. We also notice that recently, the multifactor authentication protocol has been widely concerned by researchers\cite{Far2021LAPTAS}, but these schemes are designed for authentication between users and sensors. We only focuses on the key establishment between sensor nodes in this paper. As Sciancalepore\cite{Sciancalepore2017Public} and other scholars have considered, we also believe that it is necessary to establish the authenticated session key directly between sensor nodes, especially in industrial and military applications. Sciancalepore et al. have done a lot of work in this aspect, but their scheme\cite{Sciancalepore2017Public} has no forward secrecy. Of particular relevance to our work are LSSL and NZMA, but both of them still exist some flaws mentioned above and do not consider the reliability as Watro et al. \cite{Watro2004TinyPK} do. %Futhermore, without the authentication of private keys and secret values, security is also greatly reduced. Certainly, this may be a good choice for the applications that have less demanding of security and stability, as it saves a lot of energy for nodes.
In addition, in the work described above, only the payload of EPKI and NZMA' agreement packet can be embedded into the frame conforming to IEEE 802.15.4, but they are at the expense of compromised security. 
\section{Preliminaries}\label{ch:Pre}
\subsection{Notations}
According to \cite{Law2003An}, we define the notations used in this paper as listed in Table~\ref{tab:nota}.
\begin{table}[!t]
\centering
\small
\renewcommand{\arraystretch}{1.3}
\caption{List of Notations}  
\label{tab:nota} 
  \begin{tabular}{C{1.8cm}p{15.0cm}}
    \toprule
    Notations&\multicolumn{1}{c}{Comments}\\
    \midrule
$(\cdot)_k$	& Use the key $k$ to encrypt/sign the plain text, the operation represent signing under the condition that the key $k$ is a private key.\\
$\{\cdot\}_k$ & Use the key $k$ to sign the hashed plain text\\
$\mathcal{O}$ & Infinity point on a elliptic curve $E$\\
$G$ & A generating cyclic group, used by all relevant nodes\\
$P$ & The point in the group $G$\\
$(d_A,Q_A)$	& A pair of private/public keys, $d_A$ represent for the private key of node $A$, $Q_A$ represent for the public key of $A$.\\
$N_i,N_j$ & Sensor nodes $i$ and $j$, respectively\\
$ID_i$ & The identification details or attributes of node $i$\\
$F_p$ & A prime field whose order is prime number $p$\\
$Cert_A$ & The certificate of node $A$\\
$Key_A$ & The secret value randomly created by node $A$\\
$Nonce_A$ & A temporary value for $A$ that is used to keep freshness and resist the replay attack.\\
$||$ & An operator that connects texts\\
$\oplus$	& Exclusive or operation\\
$P_{[x]},P_{[y]}$ & The $x$-coordinate and $y$-coordinate of point $P$.\\
  \bottomrule
\end{tabular}
\end{table}
\subsection{Primitives}
%RSA has a very important feature, that is, after reasonable design, signatures can be calculated with small exponents and verification needs large exponents (or vice versa), which can make the energy consumption of attackers faster than that of legitimate nodes, and thus can resist DoS attacks to a certain extent. \cite{Arazi2007A}. 
Comparing RSA \cite{Rivest1978A}, we can use the 160-bit ECC \cite{Vasundhara1987Elliptic} key to achieve the security strength of the RSA 1024-bit key. Moreover, the short key is helpful to saving storage, computation, communication resources, and many studies \cite{Gura2004Comparing,Wander2005Energy} also have shown that the ECC mechanism is more resource-efficient than RSA. So ECC is the most suitable public key scheme for resource-constrained WSNs, and the cryptographic primitives used in this paper are ECC-based. 

For any point $P$ on an elliptic curve $E$, where $E$ is a smooth curve of the long Weierstrass form

\vspace{-0.6em}\begin{displaymath}
y^2+a_1xy+a_3y\equiv x^3+a_2x^2+a_4x+a_6,
\end{displaymath}
% or \centering{$...$,}
which $a_i\in\mathbb{F}$ \cite{Malan2004A},we can construct a set $G=\{\mathcal{O},P,2P,\cdots\}$,
where $\mathcal{O}$ is an infinity point as the additive identity. Thus the Algebraic System $<G,+>$ is a cyclic Abelian group. The problem which typically involves recovery over some Galois field $\mathbb{F}$ of $k\in\mathbb{F}$, given $k\cdot P$ (called a scalar multiplication), $P$, and $E$, is called the Elliptic Curve Discrete Logarithm Problem (\textbf{ECDLP}).
The security of ECC is just based on \textbf{ECDLP}, i.e., an attacker can only solve the ECDLP when he/she wants to find the private key $k$ given points $kP$ (the public key) and $P$. At present, ECDLP has been proved to be a computationally difficult problem for appropriate parameters \cite{Vasundhara1987Elliptic}.

According to \cite{Liu2008TinyECC} on which this paper is based, we take $\mathbb{F}=F_p, a_1=a_2=a_3=0, a_4=a, a_5=6$, where $a,b\in F_p$ are constants such that $4a^3+27b^3\neq0$, $F_p$ is a prime fields and $p$ is a large prime number. So, the elliptic curve

\vspace{-0.6em}\begin{displaymath}
y^2\equiv x^3+ax+b
\end{displaymath}
is used in this paper.

The ECC key mechanism is commonly used in three ways: ECDH, ECDSA, ECIES. ECDH is a key negotiation mechanism based on ECC and Diffie-Hellman protocol, ECDSA is a digital signature scheme based on ECC, and ECIES is an encryption / decryption mechanism based on ECC. Their computational complexities are listed in Table~\ref{tab:ccECC} (according to \cite{Vasundhara1987Elliptic}) and the meaning of the relevant operations are listed in Table~\ref{tab:cos}.
\begin{table}[!t]
 \centering
 \small
 \renewcommand{\arraystretch}{1.3}
  \caption{The Complexity of ECDH, ECDSA and ECIES}
  \label{tab:ccECC}  
\begin{tabular}{p{3cm}cp{6cm}p{6cm}}
    \toprule
   \multicolumn{1}{c}{Items} & \multicolumn{1}{c}{ECDH} & \multicolumn{1}{c}{ECDSA} & \multicolumn{1}{c}{ECIES}\\
    \midrule
\tabincell{l}{A (sign/decipher)} & \tabincell{l}{2SM} & \tabincell{l}{1SM+1Hash+1Rev+2Mod+2M} & \tabincell{l}{1SM+1MAC+1Dec+1KDF}\\
\tabincell{l}{B (verify/encipher)} & \tabincell{l}{2SM} & \tabincell{l}{2SM+1PA+1Hash+1Rev+4Mod+2M} & \tabincell{l}{2SM+1Enc+1MAC+1KDF}\\
\tabincell{l}{Total} & \tabincell{l}{4SM} & \tabincell{l}{3SM+1PA+2Hash+2Rev+6Mod+4M}
& \tabincell{l}{3SM+2MAC+1Enc+1Dec+2KDF}\\
  \bottomrule
\end{tabular}
\end{table}
\begin{table}[!t]
 \centering
 \small
\renewcommand{\arraystretch}{1.3}
  \caption{The Meaning of Cipher Operation Symbols}
  \label{tab:cos}
  \begin{tabular}{p{2cm}<{\centering}lp{2cm}<{\centering}l}
    \toprule
     \multicolumn{1}{c}{Abbr.} & \multicolumn{1}{c}{Comments} &
     \multicolumn{1}{c}{Abbr.} & \multicolumn{1}{c}{Comments}\\
    \midrule
M & Multiplication & Enc & symmetric Encryption\\
D & Division & Dec & symmetric Decryption\\
Rec & Reciprocal & Hash & Hash function operation\\
Mod & Mode operation & KDF & Pseudo-random function\\
PA & Point Addition & SM & Scalar Multiplication\\
MAC & Message Authentication Code\\
  \bottomrule
\end{tabular}
\end{table}
Previously, the above three ECC schemes have been implemented on TinyOS with a configurable library called TinyECC \cite{Liu2008TinyECC}. 
\subsection{Key Establishment}
\renewcommand{\thefootnote}{\arabic{footnote}} %after display authors, set the footnote into arabic number.
In WSNs, the first thing that sensors need to do is to establish the session keys\setcounter{footnote}{0}\footnote{In fact, session key is usually generated from pairwise key by hashing function and so on \cite{Seo2015Effective}. So here we only discuss the establishment of pairwise key. The pairwise key we're talking about is actually the session key in most existing studies. } after they were deployed. Key establishment is the process by which two or more entities establish a shared secret key for subsequently security goals such as confidentiality or data integrity. In general, there are two kinds of key establishment methods: transport and agreement \cite{Law2003An}. The key transport refers to that a key is created by two entities with secure transmission. It is composed of three phases: encryption and signature, transmission, decryption and verification. The key agreement refers to both parties establish the shared secret key with exchange secret or opened information. Usually there are not explicit key primitive operations such as encryption, signature and so on. Comparably the key agreement is more computation-effective than key transport under the same security level, but the key transport can be more communication-effective. As mentioned earlier, there are no much analysis and research on key transport protocols for WSNs, so we mainly concentrate our attention on the design of AKT (Authenticated Key Transport) protocols in this paper.
\subsection{Models}
\subsubsection{Network Model}
\begin{figure}[!t]
%\begin{figure*}
%\includegraphics[width=\textwidth]{RAMsize.png}
\centering
\includegraphics[width=4.5in,clip,keepaspectratio]{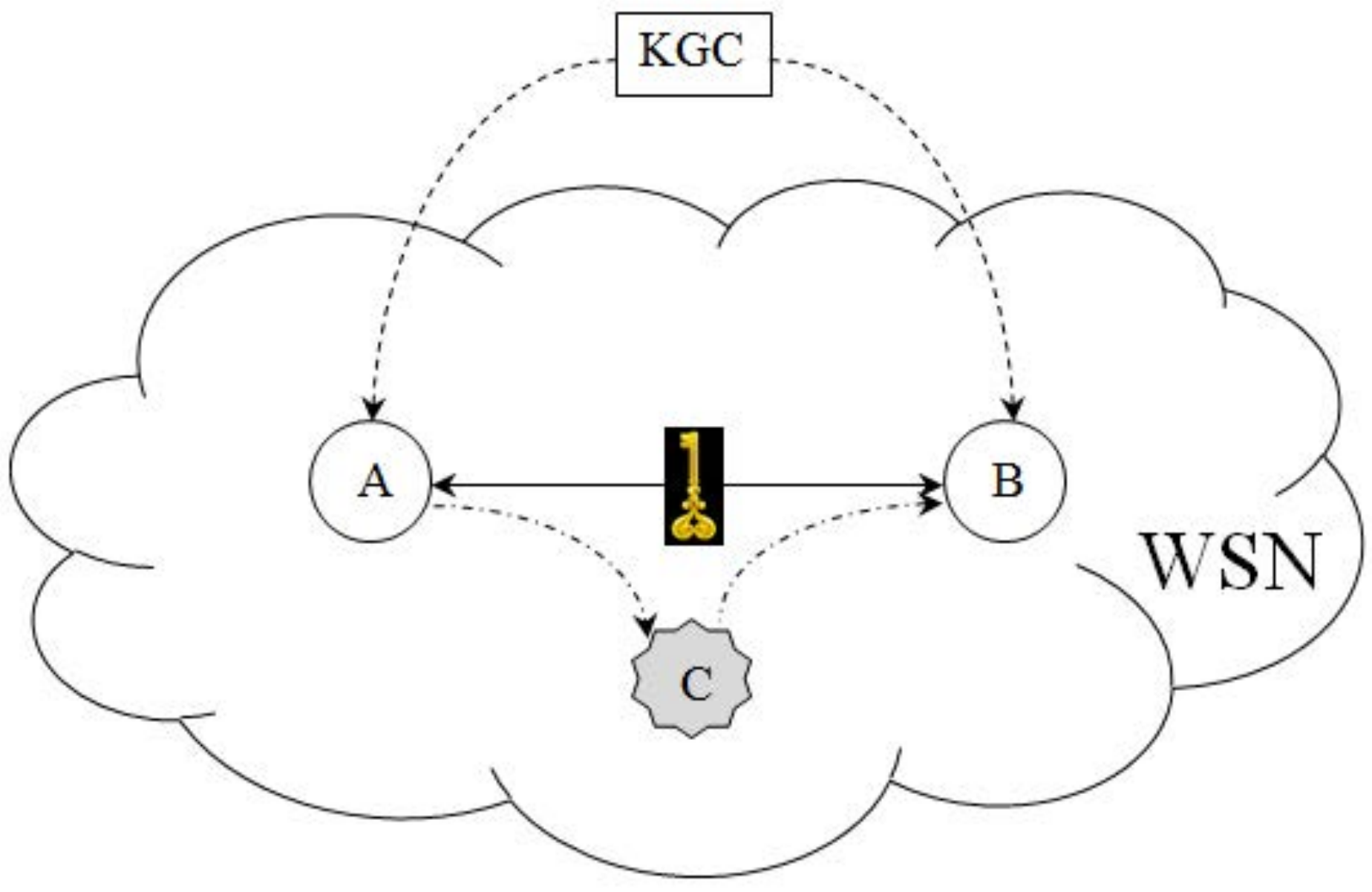}
\caption{Network Model}
\label{fig:Network}
\end{figure}

The network here we are discussing is a distributed wireless sensor network (DWSN). There are three kinds of roles in the network system. One is an off-line KGC (Key Generation Center), another is sensor nodes connected to the network, and the other is attackers (as shown in Fig. \ref{fig:Network}). KGC is in charge of the generating and loading the initial security data (including key pairs of node, public key of KGC and security parameters etc.) for every sensor node. The shared key should be established between two valid nodes (such as A, B). The attacker (e.g. $C$) has the capabilities defined by the threat model in section \ref{sec:TM}.

\subsubsection{Assumptions}
We assume that the network runtime is divided into many stages according to the renewal cycle of session key. Each stage consists of two phases: session key establishment and session key usage, and the session key usage period is usually longer than the establishment period. Thus the time of a phase is usually more than twice the time of session key establishment.% in general one stage is more than twice as long as session key establishment time. 

The correctness and security of TinyAKE are based on the following assumptions (called \textbf{Three-Assumption}), which may also be suitable for other key establishment protocols:

$\bullet$ Assumption I: The private key $d_{KGC}$ of the key generation center $KGC$ is secret. % We can describe it in a formal definition based on indistinguishability as follows:

$\bullet$ Assumption II: The private key $d_{A}$ of the sensor node $A$ is secret.

$\bullet$ Assumption III: The symmetric and asymmetric cryptographic primitive algorithms (such as encryption, digital signature) adopted in the protocol is secure. 

Formally speaking in theory of indistinguishability, encryption algorithm is secure means the algorithm can achieve $(t,\varepsilon)$-IND-CCA2 security \cite{Seo2015Effective}, that is to say, the advantage $\varepsilon$ of an attacker $A$ winning in polynomial time $t$ in the following IND-CCA2 game is negligible, i.e., $\varepsilon_{IND-CCA2}^{Adv}\leqslant negl(n)$ \cite{Bellare1993Random}, where $1^n$ is the security parameter.

Attacker $A$ sends two pieces of plaintext $M_1$ and $M_2$ to Challenger $B$, and then $B$ chooses a $M_i$ ($i$ is chosen randomly from 1,2 by $B$) to encrypt and sends it to $A$. Then $A$ guesses whether $B$ encrypted $M_1$ or $M_2$. If we represent the guess of $A$ as $M_{i}'$ ($i'=1$ or $2$) and the probability of $A$'s guess as $Pr [i'= i]$, the advantage that $A$ can win in this game is
%\vspace{-0.6em}
%\begin{displaymath}
$\varepsilon_{IND-CCA2}^{Adv}= |Pr[i'= i] - \frac{1}{2}|$.
%\end{displaymath}
And the signature algorithm is secure means the algorithm can achieve $(t,\varepsilon)$-EUF-ACMA (Existentially UnForgeable - Adaptive Chosen-Message Attack) security \cite{Seo2015Effective}, that is to say, the advantage $\varepsilon$ of attacker $A$ can forge a signature in polynomial time $t$ for his/her chosen some message is negligible, i.e., $\varepsilon_{EUF-ACMA}^{Adv}\leqslant negl(n)$.

In addition to the above assumptions, an adversary can carry out any type of attack. But if perfect forward secrecy is to be achieved, we need to assume that the attacker cannot store the packets passed in the historical phase.

\subsubsection{Threat Model}\label{sec:TM}
The attacker can control the whole communications and can eavesdrop, inject, distort, cease, forge any packets in the network. 
Here, for the sake of discussion, we classify WSN attacks into three categories (called \textbf{Three-Attack}):

$\bullet$ Attack I (node-capture attack): This kind of attack occurs when one or more nodes are captured.

$\bullet$ Attack II (packet-capture attack): Such attacks occur when one or more packets are caught.

$\bullet$ Attack III (no-capture attack): Such attacks occur without any capture, mainly include jamming, known protocol attack et al.

Obviously, the constructing difficulty of the attacks above are I~\textgreater~II~\textgreater~III.

\section{Design of AKE}\label{ch:DAKE}  % ch:PT
\subsection{Design Principles}
To better design the security schemes for WSNs, we proposed the following design principles.

\subsubsection{High Security} Although the security mechanisms can be applied according to the application requirements, our design goals should be the highest security under reasonable resource consumption. Of course, we can design the security into several levels so that users can choose according to their requirements on safety intensity. Fundamentally the confidentiality, integrity, access control (mainly authenticity) are necessary for a secure mechanism. As for non-repudiation or unforgeability, in WSNs, can be designed for applications with high security requirements, such as authenticated broadcast, confidential data uploading, et al.
As a key establishment protocol, to get better security, it should hold the following three authentications (called \textbf{Three-Authentication}):

$\bullet$ Authentication I (Public Key Authentication) \cite{Law2003An,Du2005An}: Public key is the security base of subsequent information transmission, as we know, the confidentiality is achieved by public key. But how do we know this public key belongs to our legal node, not the forger? Currently, there are two asymmetric methods for verifying the public key, one is certificate mechanism, and the other is certificateless or self-certified mechanism.

$\bullet$ Authentication II (Private Key Authentication):  we can also call it node authentication when it is still not captured. Through this authentication, we can confirm that the private key does belong to our legal node, not the attacker.

$\bullet$ Authentication III (Secret Values Authentication): Secret values for generating the pairwise key is authenticated (signed by the sender and verified by the receiver). Obviously, it is based on Authentication I and II.

%$\ \,\,$\textbf{Strong Authentication III} (Packet Authentication): every agreement packet of the protocol should be signed by the sender and verified by the receiver so that an attacker can't fake it. Obviously, Strong Authentication III implies Authentication III, since the secret values are transmitted in packet.

We refer to the key establishment protocols with Three-Authentication as AKE (Authenticated Key Establishment) protocols, the key agreement protocols with Three-Authentication as AKA (Authenticated Key Agreement) protocols and the key transport protocols with Three-Authentication as AKT  protocols. As a AKE protocol, it should make any pair of nodes achieve mutual authentication \cite{Diffie1992Authentication}.

\subsubsection{Low Power Consumption} In theory, the total energy consumption should be minimized, but it should be based on the situation to control. In the case of only symmetric key mechanism, we should reduce the communication traffic as much as possible. When the public key mechanism is introduced, we should make a compromise between communication and computation while ensuring reasonable security. Certainly, no matter which security mechanism is selected, we can make traffic get to minimum by the following methods:

$\bullet$ Minimum number of packets: We should design the protocol to minimize the number of packets under the existing standards. That is, we need to load the security data into one packet as much as possible without exceeding the maximum allowed payload of the packet. Obviously, in order to achieve mutual authentication, the minimum number of packets should be no less than 2.

$\bullet$ Minimum length of packet: After designing the protocol to minimize the number of packages, each package should be only loaded with the necessary security data to minimize the length of packet.

$\bullet$ Using broadcast as many as possible: Never use unicast when we need to notify all neighbors and can notify neighbors by broadcast.

$\bullet$ To follow the packet length limitation of IEEE802.15.4: As mentioned earlier, we'd better limit the length of packet to 127B (according to IEEE802.15.4), which is mainly for the following two considerations. One is that it is easier to integrate the hardware with software products while we meet the standard since it is adopted by many protocol stacks. The other is that it is not appropriate to use too big packet for transmitting, because we need to control the packet loss rate on the acceptable range under the bad wireless environment.

\subsubsection{Easy to Apply} The protocol should be simple, convenient to deploy, easy to extend, low computation and storage requirements. History has proved that the protocol with complex designing is often hard to put into practice.

\subsubsection{Cross-layer Designing} As mentioned earlier, we should stride across the transport layer and the network layer to design the key establishment protocol. It is necessary to consider reliability while designing protocols for key establishment, and it is more convenient to guarantee the reliability when designing key establishment protocol from the MAC layer.

\subsubsection{Scalability} First of all, to get a better scalability, the protocol designed should confirm to the existing WSN standards, e.g. IEEE802.15.4. What is more, we should make the protocol not rely on the scale of network. It is preferable that the protocol is neither dependent on the number of network nodes nor the network density. Last but not least, we should give due consideration to the scalability of platforms and software libraries. Although these issues are only considered when the protocol is implemented, the design considerations are more conducive to the smooth implementation of the protocol.
\subsection{TinyAKE}
So far, we can start the design of our protocol. We will see, the certificate of our protocol is tiny and the KGC is also tiny, the entire protocol is implemented in a tiny operating system TinyOS, and our key establishment protocol is serving the network consisting of thousands of tiny sensors. So we call the proposed protocol TinyAKE.
\subsubsection{\textbf{The Details of TinyAKE}}\quad

TinyAKE consists of two parts: TinyKGC (Tiny Key Generation Center) and MainAKE (Main phase of Authenticated Key Establishment), respectively corresponding to the Pre-distribution Stage before the deployment and the Key Establishment Stage after the deployment.

$\blacktriangleright$ \textbf{Pre-distribution Stage}: There are mainly two operations, one is to generate the initial security data, and the other is to load security data into the nodes. They are described as follows.

\textit{Generating operation}: Use the TinyKGC to generate the key pairs of the whole network, including the private/public key pair $(d_S,Q_S)$ of the server and that of every node. In the same time, the certificate of every node should be generated. The content of the tiny certificate can be defined as

\vspace{-0.6em}\begin{displaymath}
(ID,Q,\{ID||Q\}_{d_S})~~or~~(ID,Q_{[x]},\{ID||Q_{[x]}\}_{d_S})~,
\end{displaymath}
the latter is in a compressed mode, i.e. the public key $Q$ can be stored in a compressed form, which saves half the storage and communication overheads \cite{Seo2015Effective}. But the saving is exchanged with a huge amount of computation, we should consider carefully about employing it. 

\textit{Loading operation}: In this step, the security data, such as the node's own private key, certificate with public key and signed by KGC, public key of KGC, etc., will be burned into each node together with the compiled program.

\begin{figure}[!t]
\centering
\includegraphics[width=6in,clip,keepaspectratio]{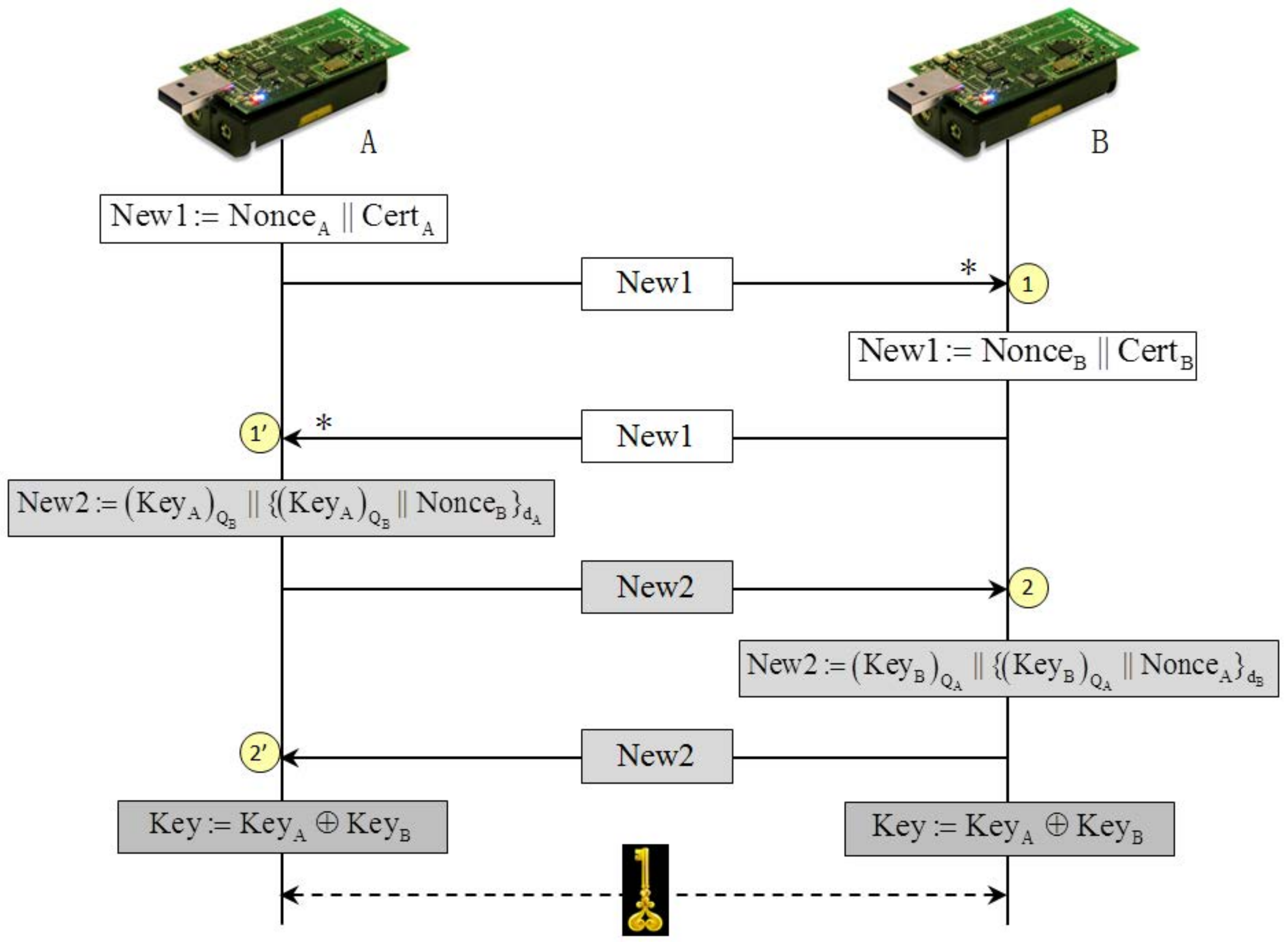}
\parbox[t]{\columnwidth}{Step 1 or 1': Upon receipt of the negotiation packet New1, it is  necessary to verify the certificate and register the security information and Nonce value when the verification is passed; Step 2 or 2': Upon receipt of the negotiation packet New2, the signature will be verified, and the pairwise key will be calculated and stored when passed.}
\caption{Key Establishment\textemdash TinyAKE}
\label{fig:TinyAKE}
\end{figure}

$\blacktriangleright$ \textbf{Key Establishment Stage}: The key establishment stage is started after the node filled with secure materials has been deployed. We have designed the key establishment protocol shown in Figure~\ref{fig:TinyAKE}. Here we define the packet New1 and New2 of node $A$ as follows, the same as other nodes:

\vspace{-0.6em}\begin{displaymath}
\begin{aligned}
&A\rightarrow\,*: New1:=Nonce_A~||~Cert_A\,,\\
&A\rightarrow{B}: New2:=(Key_A)_{Q_B}~||~\{(Key_A)_{Q_B}~||~Nonce_B\}_{d_A}.
\end{aligned}
\end{displaymath}
The whole process of the key establishment stage can be roughly divided into two phases, but the two phases can be alternated and partially overlapped. %, i.e., not completely seperated.
Basically every node only broadcasts its own certificate and a Nonce value (constituting packet New1) in the first phase, where the Nonce value is used to implement the Challenge-Response mechanism, thereby preventing replay attacks. In the second phase, when a node $A$ receives a packet New1 from node $B$, if $A$ did not share a key with $B$, then $A$ verify the correctness of the certificate in the packet New1. When the verification is passed, Node $A$ will register and store the public key and related information of node $B$ into the neighbor list. After that a random key is generated, encrypted and signed, at last sent back to node $B$ in a New2 packet. In the same time, the key to be sent will also be saved into the local neighbor list. Once node $A$ receives a packet New2 from node $C$, it will check whether the packet New1 with the same source (Node $C$) has been received (judged by the registered information). If received, after the verification is passed, the Nonce value in the packet New2 will be compared with the one stored earlier in the node $A$. If they are equal, Node $A$ will decrypt the random key from the New2 packet, and make it do an "exclusive or" operation with his/her own random key to constitute a pairwise key. At this point, Node $A$ has established a shared key with Node $C$. Certainly, When the New2 packet is reached earlier than the packet New1 with the same source, the New2 packet will be sent into a queue. If the corresponding New1 packet has not reached over a period of time, the New2 packet in the queue will be dropped.

Finally, there may be some nodes registered but not established pairwise keys, which will be cleared out of the neighbor list at the end of establishment phase.
\subsubsection{\textbf{Several Mechanisms}}\quad

$\blacktriangleright$ \textbf{Challenge-Response}: 
In order to resist the replay attack, we generate a random Nonce (because it is infrequent and the time synchronization is not easy, here we do not use the time value) for every link of a node. The Nonce is filled into the packet New1 to be sent to another party. In this way, sensor nodes can resist the responding packet New2 be copied to other field and replayed. 

$\blacktriangleright$ \textbf{Queue Buffering Mechanism}: 
Considering the unreliability of WSNs and the time-consuming of public key operations, it is necessary to buffer the received and ready packets. For the received packets, they should be put into a receive queue to do some public key operations like verification. And for sending packets, they should be cached in a send queue to do some public key operations like encryption. So in the implementation of TinyAKE, we create two queues to make it work comfortably.

$\blacktriangleright$ \textbf{Retransmission Mechanism}: 
We set a parameterized retransmission mechanism to increase the proportion of secure links and the number of retransmission can be set to any value by user before the node started. At the same time, when only one-way secure connection is established, the retransmit mechanism will also be activated. %In the same time, when only unidirectional security connections are established, retransmission will work as well.

$\blacktriangleright$ \textbf{Invalid Packet Cleanup}: 
In our TinyAKE, we set a regular cleaning mechanism that is used to clear invalid registered information and New2 packets. That is to say, after a certain period of time, if the registered neighbor has not sent valid information to establish the key, the registration entry will be cleared out from the neighbor list. And the same, the New2 packets cached in the queue will be cleared when the corresponding New1 packets have not reached in a fixed time.

\section{Proofs and Evaluation}\label{ch:PE}
\subsection{Proofs}\label{sec:proofs}
\begin{myDef}
\textbf{(Correctness)}: The protocol can work correctly, that is to say, in the case of unobstructed communication, two valid parties within the communication range can use it to establish a secure connection (pairwise key).
\end{myDef}
\begin{myTheo}
TinyAKE holds the Correctness.
\end{myTheo}
\begin{IEEEproof}
For any valid node $A$, after the protocol execute a run, it may be next four cases, for his/her every valid neighbor $B$:

$\centerdot$ The packet New1 and New2 from $B$ are received and verify passed: Obviously, $A$ can get a shared key with $B$. 

For a pair of neighboring nodes, e.g. $A$ and $B$, if they received the secret key from the other side, the pairwise key calculated by each node should be the same, because the random keys generated for a partner during an establishment stage are invariant. That is to say, if we denote the $key_B$ received by $A$ as $key'_B$ and the $key_A$ received by $B$ as $key'_A$, we have $key_A=key'_A$ and $key_B=key'_B$. Therefore, we have

\vspace{-1.2em}\begin{displaymath}
key_A\oplus{key'_B}=key_B\oplus{key'_A}.
\end{displaymath}

$\centerdot$ Only the packet New1 from $B$ is received correctly and verify passed: At this time, maybe New1 is a replay or the link is too bad to make New2 error. No matter what, node $A$ will register the neighbor $B$. But with our regular cleaning mechanism, the false neighbor $B$ will be cleared from the neighbor list of $A$.

$\centerdot$ Only the packet New2 from $B$ is received and verify passed: At this time, maybe New2 is a replay or the link is too bad to make New1 error. The same as above case, with our regular cleaning mechanism, the false New2 will be cleared from the Received Queue of $A$.

$\centerdot$ None of the packet from $B$ is received and verify passed: Maybe $B$ is a bit far from $A$, nothing needs to do, Node $A$ continues to run normally.

It is worth noting that there may be an unidirectional secure link. At this time, with our retransmission and regular cleaning mechanism, either two-way secure connections are established or the one-way "shared" key is deleted by the owner after a period of time.

In summary, node $A$ can run correctly. So the whole network with TinyAKE can work correctly.

Consequently, TinyAKE holds Correctness.
\end{IEEEproof}

\begin{myDef}\label{def:ValidNode}
\textbf{Valid Node}: A Valid Node is a node whose public and private keys can be authenticated (i.e. with Authentication I and II).
\end{myDef}

\begin{myDef}\label{def:AKE-Security}
\textbf{(AKE-Security)}: For any node $A$, if he/she uses a protocol establish a pairwise key with a node $B$, and he/she can confirm: 1). The node $B$ is a Valid Node of the secure WSN (\textbf{Authentication}); 2). The pairwise key is only known by $A$ and $B$ (\textbf{Confidentiality}). We call this protocol is AKE-secure. Formally speaking, if the used encryption primitive is $(t_1,\varepsilon_1)$-IND-CCA2 secure and the signature primitive is $(t_2,\varepsilon_2)$-EUF-ACMA secure, then the protocol is $(t,\varepsilon)$-AKE secure, where $t\geqslant min(t_1,t_2)$ is a sufficiently large polynomial time and $\varepsilon\leqslant max(\varepsilon_1,\varepsilon_2)$ is negligible.
\end{myDef}
\begin{myTheo}
TinyAKE holds the AKE-Security as the definition \ref{def:AKE-Security}.
\end{myTheo}
\begin{IEEEproof}
We can draw above conclusion by the following two lemmas.
\end{IEEEproof}

\begin{myLemma}
%$\bullet$
TinyAKE holds the \textbf{Authentication} property.
\end{myLemma}
\begin{IEEEproof}
According to the definition \ref{def:ValidNode}, we need to prove our protocol having achieved Authentication I and II. Factually in TinyAKE, the Authentication I has been achieved in Step 1 or 1' (shown in Figure~\ref{fig:TinyAKE}), anybody can verify the validity of the sender's certificate in New1 packet with KGC's public key.  And the Authentication II has also been achieved in Step 2 or 2' by verifying the signature. Formally speaking, the Authentication I is warranted by certificate, which is signed in signature algorithm. According to Assumption III, the advantage for existential forgery of the certificate in polynomial time $t$ is $\varepsilon_{EUF-ACMA}^{Adv}$, and by the definition \ref{def:AKE-Security}, $\varepsilon_{EUF-ACMA}^{Adv}=\varepsilon_2$ and $t=t_2$. Thus the advantage of forging Authentication II (need to successfully forge the certificate in New1 packet and the signature in New2 packet) is no more than $\varepsilon_{AuII}^{Adv}={\varepsilon_{EUF-ACMA}^{Adv}}^2<\varepsilon_{2}$, with polynomial time at least $2t_2$ (the time of 2 times signatures)$>t_2$. %As for Authentication III, we have authenticated the packet New2 with a signature, but for the packet New1, considering saving energy and there is a authentication on the certificate, and it is further authenticated in Step 2 or 2', so we don't authenticate it again.
Thus $\varepsilon_{AuII}^{Adv}$ is negligible. That is to say, TinyAKE holds the \textbf{Authentication} property.

In fact, our protocol has achieved Three-Authentication and any\,pair\, of nodes can achieve mutual\, authentication. Therefore, the whole\,network\,can\,achieve mutual authentication.
%On the same time, through above steps, any pair of nodes can ...
\end{IEEEproof}

\begin{myLemma}
% $\bullet$
TinyAKE holds the \textbf{Confidentiality}.
\end{myLemma}
\begin{IEEEproof}
According to definition \ref{def:AKE-Security}, we need to proof the pairwise key is only known by $A$ and $B$. In TinyAKE, as can be seen from the Figure~\ref{fig:TinyAKE}, attackers other than $A$ and $B$ who want to get the pairwise key  can only get it by the following two methods: 1). Impersonating a valid node and forging the shared secret value ($Key_A$ and $Key_B$) to compute the pairwise key. But as can be seen from Step 2 or 2' of TinyAKE, the shared secret value are signed with the private key of the sender. By \textbf{Assumption II} and \textbf{III}, during the polynomial time $2t_2$, the advantage of forging the shared secret value is no more than $\varepsilon_{2}$. That is to say, Authentication III is completed in Step 2 or 2'. 2). Revealing the shared secret value being transmitted. However the shared secret value ($Key_A$ and $Key_B$) for computing pairwise key has been encrypted with the public key of the receiving node and signed with the private key of the sending node. So according to \textbf{Assumption II} and \textbf{III}, i.e. the encryption primitive is $(t_1,\varepsilon_1)$-IND-CCA2 secure, we can get that TinyAKE holds Confidentiality and the advantage of an attacker to obtain the pairwise key in the polynomial time $t_3$ is $\varepsilon_3$, where $t_3=t_A+t_B>Max(t_A,t_B)\geqslant t_1$ and $\varepsilon_3=\varepsilon_{A}\cdot\varepsilon_{B}<Min(\varepsilon_{A},\varepsilon_{B})\leqslant\varepsilon_1$. Therefore, TinyAKE holds the \textbf{Confidentiality}.
\end{IEEEproof}

\subsection{Performance Evaluation}
\noindent $\blacktriangleright$ \textbf{Other Security}

%$\bullet$ \textbf{Confidentiality}: In our protocol, only pairwise keys need to be kept confidential since they are transmitted through public media. As can be seen from the Figure~\ref{fig:TinyAKE}, the shared secret value for pairwise key has been encrypted in a public key from the received party. By \textbf{Assumption II} and \textbf{III}, we have TinyAKE holds confidentiality.
%$\bullet$ \textbf{Authentication}: In TinyAKE, the Authentication I has been achieved in Step 1 or 1' (shown in Figure~\ref{fig:TinyAKE}), and the Authentication II has also been achieved in Step 2 or 2' by verifying the signature. As for Authentication III, we have authenticated the packet New2 with a signature, but for the packet New1, considering saving energy and there is a authentication on the certificate, and it is further authenticated in Step 2 or 2', so we don't authenticate it again.
%From the proof in Seciton \ref{sec:proofs}, we can conclude that  our protocol has achieved Three-Authentication and any pair of nodes can achieve mutual authentication \cite{Diffie1992Authentication}. Therefore, the whole network can achieve mutual authentication.

$\bullet$ \textbf{Integrity}: The integrity of certificates and packets have been implemented in TinyAKE with digit signature. Because the integrity of packet New1 has been implied in the certificate (the main part of New1), considering saving resource, we don't sign on the packet New1 again.

$\bullet$ \textbf{Resilience to Attacks}

With Three-Authentication, our protocol can resist to MITM (Man-in-the-Middle) attack,impersonate and packet forge attacks.
With Challenge-Response and regular cleaning mechanisms, our TinyAKE can resist to replay attack.

The hash operation has been used in TinyAKE, but it is just used to generate a digest so as to get a signature. The security of  signature only depends on the secrecy of private key, so Hash Collision does not influence the security of our protocol.

As for Three-Attack, our protocol can resist the Attack II and most of Attack III (except for jamming, we think it is a matter of being considered in the physical layer). Our TinyAKE can also limit the threat of Attack I to the node itself, i.e. capturing a node does not expose the security information of other nodes or servers in the network.

$\bullet$ \textbf{Resilience to Node Capture}

As described above, in TinyAKE, node capture will not expose any secret information of other nodes and KGC. With node capture and replica, attacker can only connect many nodes with same identity into the network, but there are many literature \cite{Tague2009Mitigation,Wen2012Detecting,Zhao2017On} about the detecting of the node capture attack to mitigate it.

In fact, with Three-Assumption, the security of TinyAKE only depends on ECDLP, as attackers can only get the private keys from TinyAKE by solving ECDLP. Owing to the computational security of ECDLP, TinyAKE also holds computational security.

\noindent $\blacktriangleright$ \textbf{Computational Complexity}

In TinyAKE, it is necessary for two parties to do the certificate verification, the verification and encryption / decryption of the packet New2, so the computational complexity per node is:

(2ECDSA.verify+2ECIES+2ECDSA)/2\\
=~(2SM+1PA+1Rev+4Mod+2M+1Hash)+(3SM+2MAC+1Enc+1Dec+2KDF)

+(3SM+1PA+2Hash+2Rev+6Mod+4M)\\
%=~8SM+2PA+3Rev+10Mod+6M+3Hash+2MAC+1Enc+1Dec+2KDF\\
$\approx$~8.25SM.

For the network with $n$ nodes and $e$ edges, the computational complexity per node with $d$ neighbors is:

($n\cdot d\cdot$ECDSA.verify+$2\cdot e\cdot$(ECIES+ECDSA))/$n$\\
= $d\cdot$ECDSA.verify+$2e\cdot$(ECIES+ECDSA)/$n$\\
= $d\cdot$(ECDSA.verify+ECIES+ECDSA).

\noindent $\blacktriangleright$ \textbf{Communication}

Assume the protocol works for ECC-160 and the length of packet attachment (header+tailer) is 13 (unit: bytes, the same as below), then the length of certificate is 82 (including 2 Bytes ID), the length of signature is 40, the maximum length of cipher text is 70 (51 if with compressed), the length of New1 is 4+86, and New2 with 70+40. 
For two parties, the total communication quantity of sending or receiving is:

2*(4+86)+2*(70+40)+4*13 =452 (Bytes).

For the network with $n$ nodes and $e$ edges, the sending quantity per node with $d$ neighbors is:

($n\cdot$(4+86+13)+$2\cdot e\cdot$(70+40+13))/$n$
=103+123$d$ (Bytes).

For the network with $n$ nodes and $e$ edges, the receiving quantity per node with $d$ neighbors is:

($n\cdot d\cdot$(4+86+13)+$2\cdot e\cdot$(70+40+13))/$n$
=226$d$ (Bytes).

\begin{table}[!t]
\small
\renewcommand{\arraystretch}{1.3}
  \caption{Evaluation Parameters}
  \label{tab:EvaPara}  
  \centering
  \begin{tabular}{p{3cm}<{\centering}p{3cm}<{\centering}p{3cm}<{\centering}p{3cm}<{\centering}}
    \toprule
    Operation & Time(s) & Power Consumption & SMs\\
    \midrule
ME(pu-1024bits) & 10.99 & 151.28mJ~/~time & 13.57\\
ME(pr-1024bits)$^*$ & 0.43 & 5.92mJ~/~time & 0.53\\
SM(160 bits) & 0.81 & 11.15 mJ~/~time & 1\\
Enc(AES-128) & 0.00902$^\#$ & 16.2 uJ~/~80bits & 0.0051\\
Dec(AES-128) & 0.01386$^\#$ & 24.9 uJ~/~80bits & 0.0078\\
Hash(SHA-1) & 0.03284$^\#$ & 59uJ~/~80bits & 0.0185\\
Send & \textemdash & 1184uJ~/~160bits & 0.1062\\
Receive & \textemdash & 572uJ~/~160bits & 0.0512\\
  \bottomrule
\end{tabular}
\parbox{416pt}{\vspace{0.3em}*:$e=2^{16}+1$, \#: Converting from~ \cite{Carman2000Constraints}, SMs: Normalize to SM}
\end{table}

\noindent $\blacktriangleright$ \textbf{Energy Consumption}
\begin{table*}[!t]
\footnotesize
\renewcommand{\arraystretch}{1.3}
 \caption{Performance Evaluation*}
  \label{tab:PerfEval}  
  \centering
  \belowcaptionskip=-1pt
  %\addtolength{\intextsep}{5pt}
  \belowrulesep=0pt
  \aboverulesep=0pt
  \belowbottomsep=3pt
  \begin{tabular}{C{1.2cm}|c|c|c|c|c|c|c|c|c}
  %C{1.8cm}|C{1.9cm}|C{1.8cm}|C{1.9cm}}
  \toprule[0.08em]
  \rule{0pt}{10pt} % heighting
    %\rowcolor{gray}{.9}
    \multirow{3}{*}{Items} & Protocols & EPKI\textsuperscript{\cite{Malan2004A}} & TinyPK\textsuperscript{\cite{Watro2004TinyPK}} & \multicolumn{2}{c|}{LSSL \textsuperscript{\cite{Wander2005Energy}}}& CL-EKM\textsuperscript{\cite{Seo2015Effective}} & NZMA \textsuperscript{\cite{Nadir2016Establishing}} & AKAI\textsuperscript{\cite{Saeed2019AKAIoTs}}& TinyAKE\\
   \cline{2-10}
\rule{0pt}{10pt}&Type&\circled{5}&\circled{1}&\circled{1}&\circled{3}&\circled{2}&\circled{4}&\circled{2}&\circled{1}\\
   \cline{2-10}
\rule{0pt}{10pt}&Primitives&ECC&RSA&RSA&ECC&ECC&ECC&ECC&ECC\\  
\midrule
\multirow{11}{*}{\tabincell{c}{Security\\(80-bit)}}&Confidentiality&Yes&Yes&Yes&Yes&Yes&Yes&Yes&Yes\\&Integrity&&Yes&Yes&&&&&Yes\\
&Authentication I&&One-way&Yes&Yes&Delayed&Yes&Yes&Yes\\
&Authentication II&&One-way&Yes&&Delayed&&Yes&Yes\\
&Authentication III&&Yes&&&Delayed&&&Yes\\
&Resis. to MITM&&One-way&Yes&&Yes&Yes&Yes&Yes\\
&Resis. Replay&&Yes&Yes&Yes&Delayed&&Yes&Yes\\
&Resis. Impersonation&&One-way&Yes&&Yes&&Yes&Yes\\
&Resis. Forging&&One-way&&&Yes&&Yes&Yes\\
&Resis. Hash Collision&&Yes&Yes&&Yes&&Yes&Yes\\
&Ephemenal Key&Yes&Yes&Yes&&Yes&Yes&Yes&Yes\\
\midrule
\multirow{4}{*}{Reliability}&Consider Lossy&&Yes&&&Yes&&&Yes\\\vspace{-3em}&Care Failed Verifing&&Yes&&&&&&Yes\\
&Confirmation&&Implicit&&&Yes&Yes&&Implicit\\
&Retransmission&&&&&Yes&&&Yes\\
\midrule
\multirow{3}{*}{Scalability}&on IEEE 802.15.4&Yes&&&&&Yes&&Yes\\&Dep. NumOfNodes&&&&&&Yes&&\\
&Dep. Net Density&Yes&Yes&Yes&Yes&Yes&Yes&Yes&Yes\\ 
\midrule%\rule{0pt}{10pt} 
\multirow{5}{*}{Expenses}&Computation  (SM)&4&41.80&55.37&6.05&10.13&6.25&12.09&16.49\\ %$\approx$
&Communication (B)&106&609&934&341&387&276&304&452\\
&Extra RAM (B)&20d&148d&148d&60d&60d&23n+20d&60d&60d\\
&PC in Comp. (mJ)&44.60&583.69&734.97&67.06&112.95&69.67&134.75&183.85\\
&PC in Comm. (mJ)&9.31&53.47&82.01&29.94&33.98&24.23&26.69&39.69\\
  \bottomrule
\end{tabular}
\parbox[t]{\textwidth}{* To get more clear, we use the space to represent for "No".  The data of computation and communication is only for two parties, as for the whole network we should reference the corresponding analysis.$\quad$PC: Power Consumption$\quad$d: Number of Neighbors$\quad$Resis.: Resistance to$\quad$Dep.: Depending on$\quad$Comm.: Communication$\quad$Comp.: Computation
Type: \circled{1}: AKT with Certificate, \circled{2}: Certificateless AKA, \circled{3}: AKA with Certificate, \circled{4}: AKT with Hash Authentication, \circled{5}: Key Agreement without Authentication
} 
\end{table*}

The energy consumption is mainly from the consumption of computation and communication. With the quality of computation and communication and the energy consumption of every unit shown in Table \ref{tab:ccECC} and Table \ref{tab:EvaPara}, we can get the consumption very easy. For the case of two parties, the energy consumption has been calculated and listed in Table \ref{tab:PerfEval}.

%\vspace{-0.6em}
\subsection{Comparison and Improvement}
In order to investigate the performance such as feasibility and security of our scheme, we compare TinyAKE with several existing related protocols under 80-bit security strength, as shown in Table \ref{tab:PerfEval}. The evaluation parameters, as listed in Table \ref{tab:EvaPara}, is mainly referred to \cite{Gura2004Comparing} and \cite{Wander2005Energy}. In addition to the earlier TinyPK and LSSL schemes, recently only NZMA is the closest to our TinyAKE. But in order to better investigate the relationship between key transport and key agreement, and the relationship between certificated and certificateless schemes, we also put two recent certificateless schemes into comparison. 

In contrast to these state-of-the-art schemes, primarily our scheme is more practicable and trustable under reasonable resource consumption. %In contrast to other schemes, primarily our scheme is more practicable and secure but not consuming too much resources.
At first, we design the agreement packet carefully so as to meet the standard IEEE802.15.4, which make the hardware and software designed for WSNs more compatible. There are only three protocols (EPKI, NZMA and our protocol TinyAKE) which make the packet length no more than the maximum allowed payload, but the security of the former two is much lower than TinyAKE.

To illustrate the problem, a small case is designed here, so that we can explain why the security of our protocol is better than other protocols. Just take the forging and MITM attack for example, in NZMA,  an attacker can listen to and copy the public key of some $B1$ in $B$ region, then impersonates himself/herself as $B1$ and comes into $A$ region. When some $A1$ in $A$ region broadcast a request to establish key with somebody, the adversary $T$ will send the public key of $B1$ to $A1$. Because the public key is valid, so it can be verified successfully by $A1$. And then $T$ reply the pairwise key which is generated by $T$ itself and encrypted with the public key of $A1$. At last, $A1$ thought he/she have shared a pairwise key with $B1$ but actually with $T$. In the same way, $T$ can establish a key with $B1$. Therefore, a MITM attack is achieved.

On the same time the security of AKE protocols are more perfectly defined in both formal and non-formal ways. And we have proved that TinyAKE holds Security as defined in definition \ref{def:AKE-Security} and has implemented Three-Authentication which can resist most non-capture attacks (Attack II and III) such as impersonating, forging, MITM and so on. Of the existing protocols, only TinyPK did so, but is designed for RSA with huge resource consumption.

As for the high expenditure, we argue, since we have chosen the public key scheme, we should make it more trustable (secure and dependable) in the case of runnable. 
%And with the continuous development of hardware and software, a marginally higher expense is acceptable \cite{Targhetta2015Energy,Nachman2008IMOTE2}. Meanwhile, with the development of rechargeable WSNs, the occasional extravagant use of electricity in sensor nodes is also permitted \cite{He2013Energy}. Therefore, although with high expense, as Wander et al. \cite{Wander2005Energy} described, our protocol will only be used at infrequent several times but can win a high security for the whole life of the network. In a word, it is valuable and practicable.
Therefore, although our TinyAKE costs a little more than other protocols, it has a higher security level, which is also necessary in some applications.  Besides, our scheme is more efficient in the same category. Due to using RSA primitive, although TinyPK and LSSL-RSA have similar security, their computation and communication traffic are over 2.5 and 1.3 times of TinyAKE respectively. 

%\vspace{-0.6em}
\section{Experiment and Analysis}\label{ch:EA}
\subsection{Experiment}
\subsubsection{Experimental Environment}
The experimental environment parameters of hardware and software are listed in Table \ref{tab:exper}. Considering that WSNs mostly use economic sensor nodes e.g. TelosB which has the support of TinyOS and TinyECC, and experiments show that using Contiki and high-configured IoT devices does not seem to get faster than the schemes based on TinyOS, we still do our experiment on TelosB based on TinyOS and TinyECC.
\begin{table}[!t]
\small
\renewcommand{\arraystretch}{1.3}
  \caption{Experimental Environment}
  \label{tab:exper}
  \centering
  \begin{tabular}{rl}
    \toprule
    \multicolumn{1}{c}{Items} & \multicolumn{1}{c}{Configurations}\\
    \midrule
Node OS & TinyOS 2.1.1\\
Primitives library & TinyECC 2.0\\
KGC Hardware & Intel i7-4790 3.6 GHz 16 G RAM\\
KGC OS & Ubuntu 12.04\\
Sensor Node & TelosB\\
Simulation Node & MICAz\\
  \bottomrule
\end{tabular}
\end{table}
\subsubsection{Test the limitation of frame length}
According to IEEE 802.15.4, the max length of a frame is 127B (aMaxPHYPacketSize) \cite{Ieee2012IEEE}, so the payload of a frame is no more than 118B after the header and tailer of the frame are removed according to reference  \cite{Wander2005Energy}. However, the maximum allowed payload is 114B in our radio CC2420 test based on TinyOS 2.1.1. Therefore, the packet of key establishment should be no bigger than 114 Bytes in TinyOS.

\subsubsection{TinyAKE Experiment}
\begin{figure}[!t]
%\begin{figure*}
%\includegraphics[width=\textwidth]{RAMsize.png}
\centering
\small
\includegraphics[width=5inc]{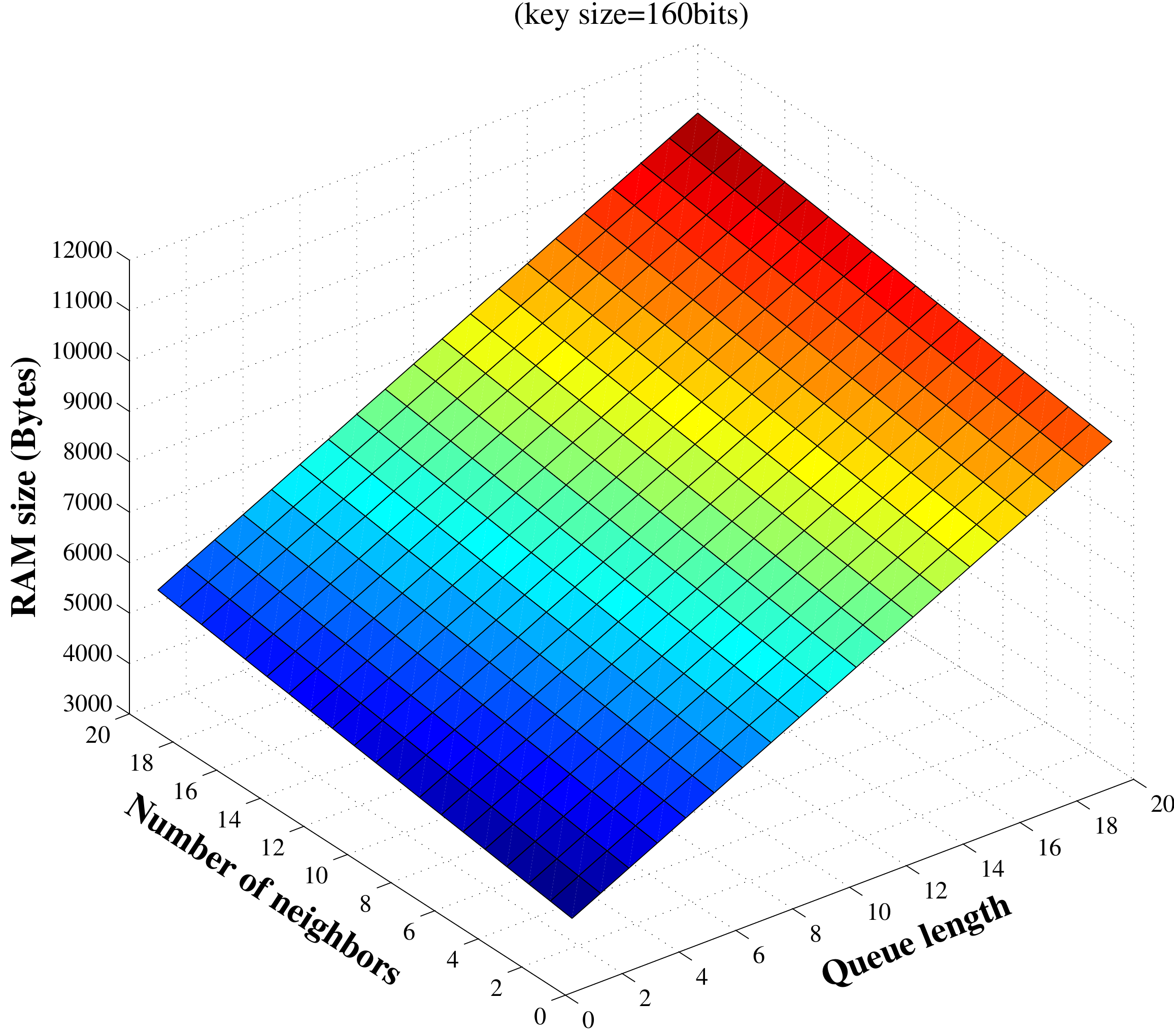}
\caption{RAM Size Increases Along with the Number of Neighbors and Queue Length}
\label{fig:RAMsize}
\end{figure}

We implement TinyAKE in an embedded programming language nesC \cite{Gay2003The} based on TinyOS. The implementation is completed with two projects. One project called TinyKGC is used for generating all the initial security data, including the certificates and public/private key pairs of all nodes and KGC itself. Another project called TinyAKE, which is loaded with initial security data from TinyKGC, is used to implement the authenticated key establishment. The total protocol is based on the primitives library TinyECC. The storage cost of some projects (ECC-160) after compiled is shown in Table~\ref{tab:RAMsize}. TinyAKE's storage cost is equivalent to NZMA.
\begin{table}[!t]
\small
\renewcommand{\arraystretch}{1.3}
  \caption{Comparing the Storage Size of Several Projects}
  \label{tab:RAMsize}
  \centering
  \begin{tabular}{ccc}
    \toprule
    Project/Protocol & RAM(Bytes) & ROM(Bytes)\\
    \midrule
TinyECC & 738 & 14140\\
TinyKGC & 2444 & 27504\\
TinyAKE & 4028 & 37890\\
NZMA & 4292 & 31066\\
EPKI & 1140 & 34342\\
  \bottomrule
\end{tabular}
\end{table}
The data of TinyAKE listed in the table is just for the case that the max number of neighbors is 1 and the queue length is 1. Obviously it's not enough to put into practice. In fact, the RAM storage size will increase with the maximum number of neighbors and the queue length. Each additional neighbor will take up 64B, and 284B storage will be used for each additional unit of queue length. The RAM size along with the number of neighbors and the queue length is shown on Figure~\ref{fig:RAMsize}. We can choose a suitable queue length and number of neighbors from it according to the actual storage of different applications. For example, when our node platform has 8KB RAM and 160 bit keys, we can set the neighbor number to 16 and the queue length to 12.
\begin{table*}[!t]
\small
\renewcommand{\arraystretch}{1.3}
\caption{Experiment Analysis*}
  \label{tab:ExpAnal}
  \centering
  \belowrulesep=0pt
  \aboverulesep=0pt
  \belowbottomsep=3pt 
  \begin{tabular}{p{1.2cm}|C{1.4cm}|c|c|C{1.1cm}|c|C{2cm}|c|c|C{1.2cm}}%<{\centering}
  \toprule[0.08em]
  \rule{0pt}{10pt} % heighting
    %\rowcolor{gray}{.9}
    \multirow{2}{*}{Items} & Protocols & EPKI\textsuperscript{\cite{Malan2004A}} & TinyPK\textsuperscript{\cite{Watro2004TinyPK}} & \multicolumn{2}{c|}{LSSL\textsuperscript{\cite{Wander2005Energy}}} & CL-EKM\textsuperscript{\cite{Seo2015Effective}} & NZMA\textsuperscript{\cite{Nadir2016Establishing}} & AKAI\textsuperscript{\cite{Saeed2019AKAIoTs}}& TinyAKE\\
   \cline{2-10}
\rule{0pt}{10pt}&Primitives&ECC&RSA&RSA&ECC&ECC&ECC&ECC&ECC\\

\midrule
\multirow{4}{*}{Expenses}&\rule{0pt}{10pt} ET (s)&34.173&481&110.741&12.101&31.514&21-22&38.047&66.5\\
&NET (s)&31.537&443.891&55.370&6.051&63.028&21-22&152.188&66.5\\
&RAM(B)&1140&1167&n/a&n/a&n/a&4292&4758&4028\\
&ROM(B)&34342&12408&n/a&n/a&n/a&31066&47236&37890\\

\midrule %\rule{0pt}{10pt}
\multirow{3}{*}{Platform}&Hardware&MICA2&MICA2&\multicolumn{2}{c|}{MICA2Dots}&TI EXP5438&TelosB&Z1&TelosB\\
&TinyOS&1.x&1.x&\multicolumn{2}{c|}{1.x}&Contiki&  1.1.10&Contiki&2.1.1\\
&TinyECC&n/a&n/a&\multicolumn{2}{c|}{n/a}&2.0&1.0&RELIC&2.0\\
  \bottomrule
\end{tabular}
\parbox[t]{\textwidth}{* The data of computation and communication is only for two parties, as for the whole network we should reference the corresponding analysis. ET: Establishment Time, NET: Normalized Establishment Time.
%Hardware: \textcircled{1}: MICA2, \textcircled{2}: MICA2Dots, \textcircled{3}: TelosB Power Consumption: P1/P2, where P1 means the power consumption on computations and P2 means communications.
}
\end{table*}
At last, the key establishment experiment is finished on two TelosB nodes, the average establishing time is about 66.5s for ECC-160 and 62.2s for ECC-128.
\begin{figure}[!t]
\centering
\setlength{\abovecaptionskip}{0.3cm}
\includegraphics[width=4in]{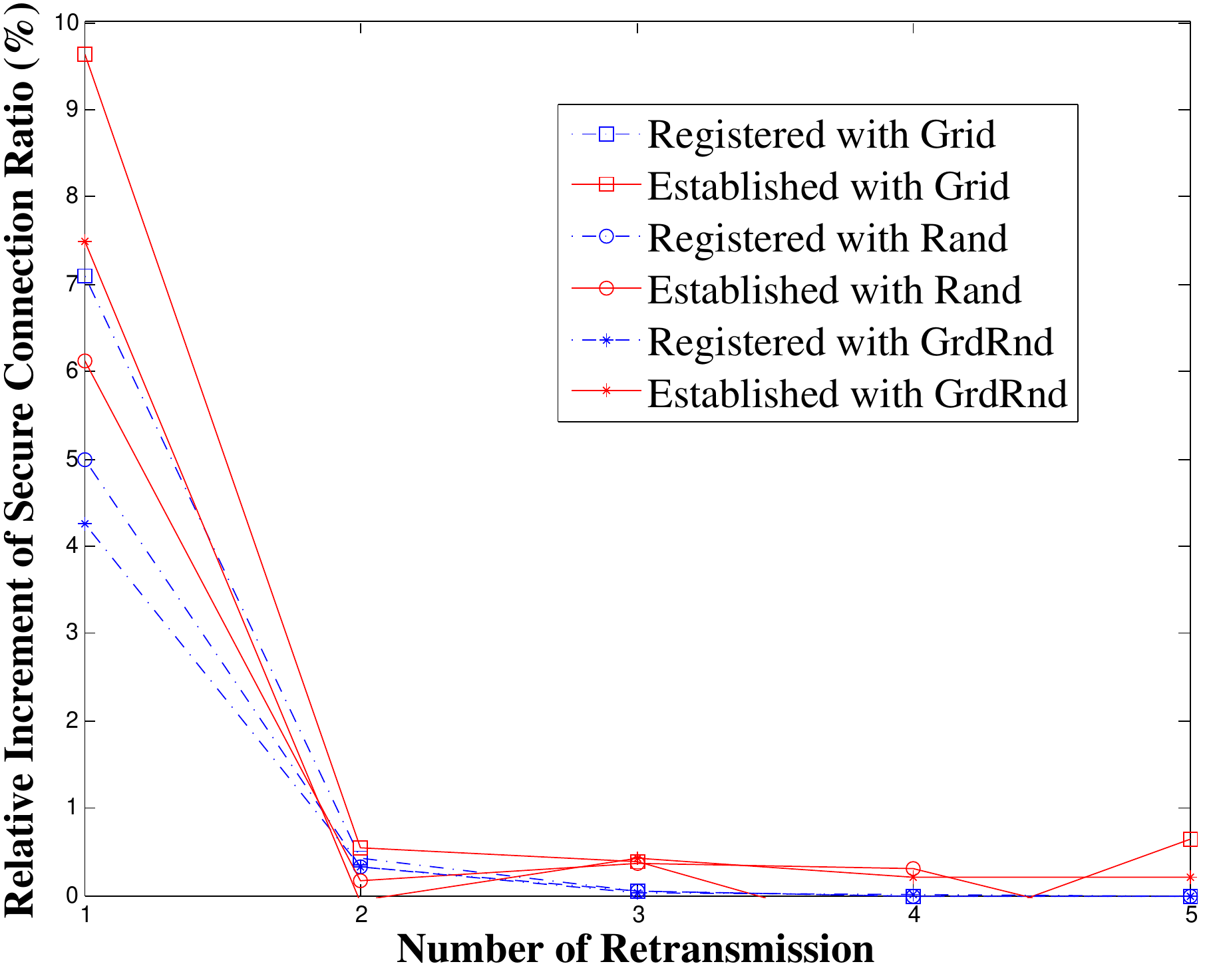}
\caption{\small{Secure Connection Ratio and Retransmissions}}
\label{fig:Retrans}
\end{figure}
\subsubsection{Simulation on Retransmission}
%\begin{anonsuppress}
According to the work of Sun et al. % \cite{Sun2009Repetitious}
, repeated negotiation is helpful to improve the security connection rate. 
%\end{anonsuppress}
In order to study the impact of repeated broadcast New1 packets on the establishment of secure links, we have carried out the simulation study of 6,49, and 961 nodes in TOSSIM. Here we take 961 nodes as an example. The topology of 961 nodes is arranged into three layouts: grid, uniform, and random. 
%The topologies of these three layouts are displayed in Figure~\ref{fig:Topo}.
All of the nodes are deployed in a field of 750mx750m. In the grid layout, there is a 30x30 mesh grid on the field, nodes are arranged at the intersection of grid dividing lines (including the sidelines) and the distance between nodes is 25m. In the uniform layout, there is a 31x31 mesh grid on the field, 961 nodes are evenly allocated into 961 grids, and then each node is placed in the grid randomly. While in the random layout, all the nodes are randomly distributed into the field. The file "casino-lab.txt" of TinyOS is adopted as the noise model file.
%\begin{figure*}
%\setlength{\abovecaptionskip}{0.3cm}
%%\setlength{\belowcaptionskip}{0.cm}
%\includegraphics[width=\textwidth]{figs/Topology.pdf}
%\caption{Three Topologies and Layouts}
%\label{fig:Topo}
%\end{figure*}
To find the optimal value of retransmissions, we designed 6 groups of experiments according to the number of repetitions. 10 experiments were repeated in each group, and the average was obtained after the results of each group were recorded. The final result is listed in Figure~\ref{fig:Retrans}. As we can see from Figure~\ref{fig:Retrans}, the first retransmission of the packet New1 lead to a 9.64\% relative increase in secure connection ratio. But the effect of the second retransmission is not obvious, just 0.55\%, and it can be ignored relative to its cost. Therefore, we think the 1 retransmission (or 2 rounds key establishment) is suitable for TinyAKE, and it is referable value for the other public-key-based key establishment protocols.
%\vspace{-2.6em}
\subsection{Analysis}
We have compared experimental results of TinyAKE with the existing representatives of various protocols as shown in Table \ref{tab:ExpAnal}. Compared with the similar schemes, before the test values are normalized to an unified platform (8bits \& 8MHz), our protocol seems to be slower than other protocols. But our protocol is running on TelosB, a platform with 16bits \& 4MHz MCU, and CL-EKM is running on TI EXP5438, a platform with 16bits \& 8MHz MCU, AKAI is running on Z1, platform with 16bits \& 16MHz MCU. Therefore, the normalized establishment time of our scheme is equivalent to that of CL-EKM and is less than $\frac{1}{2}$ that of AKAI. Our scheme is slightly more expensive than CL-EKM, but the delay authentication of our scheme is shorter than that of CL-EKM, moreover the packet length of our protocol is under the limitation of IEEE802.15.4's max frame length. 
%more scalable (as shown in Table \ref{tab:PerfEval} and \ref{tab:ExpAnal}). In the meanwhile,
In terms of compliance with existing standards e.g. IEEE802.15.4, our scheme is the most effective among the existing standards-compliant schemes.

\section{Conclusions}\label{ch:Con}
Using public key to establish symmetric session key is a classical method,  whose application on WSNs has been paid more attention in the past more than 10 years. But prior protocols are not enough practicable, since their agreement packets are beyond the maximum frame length and the need of multi-round key establishment has not been considered. Most importantly, most protocols have security flaws. To address these problems, a set design rules of WSN security are proposed. And then, according to these rules, TinyAKE, a ECC-based key establishment protocol employing the certificate mechanism and key transport mode, is presented. The correctness and security of TinyAKE are proved and analyzed. In the mean while its performance is evaluated and compared with the existing protocols. Finally the protocol is implemented in the TinyOS with library TinyECC. Experiment on TelosB and simulation based on MICAz are completed. The experimental results show that the key transport with certificate mechanism is feasible in WSNs. The simulation results show that it is suitable for the key agreement to be done with 2 rounds (including 1 repeated). The final evaluation shows that TinyAKE is more practical and trustable than existing protocols.

In addition, we propose a cross-layer design idea and some security concepts such as Three-Authentication, Three-Attack and Three-Assumption to discuss and analyze security problems. But how to apply these ideas to build a securer and more reliable public key infrastructure for WSNs will be our further work.
%In the same time, we propose a cross-layer design idea to design a key establishment protocol from the data link layer or even the physical layer (across the transport and network layers), we also present some security concepts such as Three-Authentication, Three-Attack and Three-Assumption to discuss and analyze security problems. But how to apply these ideas to build a securer and more reliable public key infrastructure for WSNs will be what we are going to study further.


\section*{Funding Statement}
This work was supported by the National Natural Science Foundation of China [grant numbers: 61972293, 61502346 and 61902189]. This work was supported in part by the Research Foundation of Education Bureau of Hunan Province, China under [grant numbers: 19B450].

\section*{Acknowledgment}
The authors would like to thank all the reviewers for their insightful comments and kind guidances to improve this paper. And also thanks to these free platforms such as  TinyECC2.0 and TinyOS.
\ifCLASSOPTIONcaptionsoff
  \newpage
\fi



%\bibliographystyle{IEEEtran}

% %\begin{thebibliography}{1}
\bibliographystyle{IEEEtran} %unsrt}
\bibliography{IEEEabrv,sample-bibliography} 
% %\bibitem{IEEEhowto:kopka}
% %H.~Kopka and P.~W. Daly, \emph{A Guide to \LaTeX}, 3rd~ed.\hskip 1em plus
% %  0.5em minus 0.4em\relax Harlow, England: Addison-Wesley, 1999.

% %\end{thebibliography}

%\balance
%\vfill


\end{document}